\documentclass{aip-cp}

\usepackage[numbers]{natbib}
\usepackage{graphicx}

\begin{document}

\title{Plasma Effects On Atomic Data For The K-Vacancy States Of Highly Charged Iron Ions}

\author[aff1]{J. Deprince}
\author[aff2,aff3]{S. Fritzsche}
\author[aff4]{T. Kallman}
\author[aff1]{P. Palmeri}
\author[aff1,aff5]{P. Quinet\corref{cor1}}

\affil[aff1]{Physique Atomique et Astrophysique, Universit\'e de Mons, B-7000 Mons, Belgium.}
\affil[aff2]{Helmholtz Institut Jena, 07743 Jena, Germany.}
\affil[aff3]{Theoretisch Physikalisches Institut, Friedrich Schiller Universit\"at Jena, 07743 Jena, Germany.}
\affil[aff4]{NASA Goddard Space Flight Center, Code 662, Greenbelt, MD, USA.}
\affil[aff5]{IPNAS, Universit\'e de Li\`ege, Sart Tilman, B-4000 Li\`ege, Belgium.}
\corresp[cor1]{Corresponding author: Pascal.Quinet@umons.ac.be}

\maketitle

\begin{abstract}
The main goal of the present work is to estimate the effects of plasma environment on the atomic parameters associated with the K-vacancy states in highly charged iron ions within the astrophysical context of accretion disks around black holes. In order to do this, multiconfiguration Dirac-Fock computations have been carried out by considering a time averaged Debye-H\"uckel potential for both the electron-nucleus and electron-electron interactions. In the present paper, a first sample of results related to the ionization potentials, the K-thresholds, the transition energies and the radiative emission rates is reported for the ions Fe$^{23+}$ and Fe$^{24+}$.
\end{abstract}

\section{INTRODUCTION}
X-ray emission lines from accreting sources, most notably the K$_{\alpha}$ and K$_{\beta}$ lines from iron ions, have observed widths and shifts which imply an origin very close to the compact object in many cases [1]. The inferred line origin can be near either the innermost stable circular orbit or the event horizon in the case of a black hole. The intensity of these lines can provide insight into the amount of gas and other properties, including the effects of special and general relativity in the emitting region, and this information is not available from other observational techniques.

Much of what we can learn from these K$_{\alpha}$ and K$_{\beta}$ emission lines depends on the use of reliable atomic parameters that allow us to infer the rate at which ions emit or absorb in line transitions under various conditions. In the case of iron ions, for example, these atomic parameters allow us to derive the number of iron ions responsible for line emission observed from different objects, which in turn gives information on the fractional abundance of iron relative to other elements (e.g. hydrogen). However, the rates and assumptions employed in these model calculations are all based on isolated iron ions. They do not account for the true situation which is a dense plasma in which the effects of nearby ions and electrons can have significant effects on the processes affecting line emission and the survival or destruction of iron ions. Although dynamical models for black hole accretion flows appear to support the existence of rather high densities, up to 10$^{20}$ - 10$^{21}$ cm$^{-3}$ [2,3], their effect on line emission has not been explored so far. It is worth noting that such high densities are required in order to allow the survival of iron ions against ionization near a black hole.

The main goal of the present work is to estimate the effects of plasma environment on the atomic parameters associated with the K-vacancy states in highly charged iron ions. In order to do this, multiconfiguration Dirac-Fock computations have been carried out for these ions by considering a time averaged Debye-H\"uckel potential for both the electron-nucleus and electron-electron interactions using a combination of the GRASP92 code [4] for obtaining the wavefunctions and the RATIP code [5] for computing the atomic parameters. A first set of results related to the ionization potentials, the K-thresholds, the transition energies and the radiative emission rates for two particular highly charged iron ions, i.e. He-like Fe XXV and Li-like Fe XXIV, is reported in the present paper.

%

\section{THEORETICAL APPROACH}

\subsection{Relativistic Multiconfiguration Dirac-Fock Method}

The wavefunctions in Fe XXIV and Fe XXV ions were obtained using the fully relativistic multiconfiguration Dirac-Fock (MCDF) method with the GRASP92 version [4] of the General-purpose Relativistic Atomic Structure Program (GRASP) initially written by Grant and coworkers [6-8]. In this approach, the atomic state functions (ASF), $\Psi$($\gamma$$J$$M_J$), are expanded in linear combinations of configuration state functions (CSF), $\Phi$($\alpha$$_i$$J$$M_J$), according to

\begin{eqnarray}
\Psi(\gamma J M_J) = \sum_i c_i \Phi (\alpha_i J M_J)
\end{eqnarray}

The CSF are in turn linear combinations of Slater determinants constructed from monoelectronic spin-orbitals of the form :

\begin{eqnarray}
\varphi_{n \kappa m} (r, \theta, \phi) = \frac{1}{r} \left( \begin{array}{c} P_{n \kappa}(r) \chi_{\kappa m} (\theta,\phi) \\ i Q_{n \kappa}(r) \chi_{- \kappa m} (\theta,\phi) \end{array} \right)
\end{eqnarray}

where $P_{n \kappa}$($r$) and $Q_{n \kappa}$($r$) are, respectively, the large and the small component of the radial wavefunction, and the angular functions $\chi$$_{\kappa m}$($\theta$,$\phi$) are the spinor spherical harmonics. The $\alpha$$_i$ represent all the one-electron and intermediate quantum numbers needed to completely define the CSF. $\gamma$ is usually chosen as the $\alpha$$_i$ corresponding to the CSF with the largest weight $|$$c_i$$|$$^2$. The quantum number $\kappa$ is given by $\kappa$ = $\pm$ ($j$+1/2) where $j$ is the quantum number associated to the total kinetic moment of the electron. The radial functions $P$$_{n \kappa}$($r$) and $Q$$_{n \kappa}$($r$) are numerically represented on a logarithmic grid and are required to be orthonormal within each $\kappa$ symetry. In the MCDF variational procedure, the radial functions and the expansion coefficients $c_i$ are optimized to self-consistency.

In both Fe XXIV and Fe XXV ions, we considered the active space (AS) method for building the MCDF multiconfiguration expansion, the latter being produced by exciting the electrons from the reference configurations to a given set of orbitals. More precisely, for Fe XXIV, the AS contained all the single and double excitations from the (1s+2s+2p)$^3$ reference configurations to the $n$=3 orbitals.  This gave rise to the following set of 44 electronic configurations : 1s$^2$(2s, 2p, 3s, 3p, 3d), 1s2s(2p, 3s, 3p, 3d), 1s2p(3s, 3p, 3d), 1s(2s$^2$, 2p$^2$, 3s$^2$, 3s3p, 3s3d, 3p$^2$, 3p3d, 3d$^2$), 2s$^2$(2p, 3s, 3p, 3d), 2s2p(3s, 3p, 3d), 2s(2p$^2$, 3s$^2$, 3s3p, 3s3d, 3p$^2$, 3p3d, 3d$^2$), 2p$^2$(3s, 3p, 3d), 2p(3s$^2$, 3s3p, 3s3d, 3p$^2$, 3p3d, 3d$^2$) and 2p$^3$. In the case of Fe XXV, all the single and double excitations from the 1s$^2$ ground configuration to the $n$=2 and $n$=3 orbitals were considered in the MCDF calculations. This led to the following multiconfiguration expansion : 1s$^2$, 1s(2s, 2p, 3s, 3p, 3d), 2s$^2$, 2s(2p, 3s, 3p, 3d), 2p$^2$, 2p(3s, 3p, 3d), 3s$^2$, 3s(3p, 3d), 3p$^2$, 3p3d and 3d$^2$. The computations were done with the extended average level (EAL) option, optimizing a weighted trace of the Hamiltonian using level weights proportional to 2$J$+1, and they were completed with the inclusion of the relativistic two-body Breit interaction and the quantum electrodynamic corrections (QED) due to self-energy and vacuum polarization.

The MCDF ionic bound states generated by the GRASP92 code were then used in the RATIP program [5] to compute the atomic structure and radiative parameters associated with the K-vacancy states.

\subsection{Debye-H\"uckel Model Potential}
In order to model the effects of the plasma screening on the atomic properties of the highly charged iron ions, we used a Debye-H\"uckel potential, which, in atomic units (a.u.) reads as

\begin{eqnarray}
V^{DH}(r,\lambda)=-\sum_{i=1}^{N} \frac{Ze^{-\lambda r_{i}}}{r_{i}} + \sum_{i>j}^{N} \frac{e^{-\lambda r_{ij}}}{r_{ij}}
\end{eqnarray}

where $N$ is the number of bound electrons, $r_i$ is the distance of the $i$th electron from the nucleus, and $r_{ij}$ is the distance between the electrons $i$ and $j$. Moreover, the plasma screening parameter, $\lambda$, is the inverse of the Debye shielding length, $\lambda$$_{De}$, and can be expressed in terms of the electron density, $n_e$, and temperature, $T_e$, of the plasma as

\begin{eqnarray}
\lambda = \frac{1}{\lambda_{De}} = \left(\frac{4 \pi n_e}{k T_e}\right)^{1/2} (a.u.)
\end{eqnarray}

Any given value of this latter parameter is obviously associated with a certain plasma environment. As an example, according to the magnetohydrodynamics (MHD) simulations reported by Schnittman {\it et al} [3] for accreting black holes with ten solar masses and an accretion rate of 10\%, the plasma parameters should be $T_e$ = 10$^5$ -- 10$^7$ K and $n_e$ = 10$^{18}$ -- 10$^{21}$ cm$^{-3}$. This corresponds to values up to about 0.1 a.u. for the screening parameter $\lambda$, as shown in Table 1. Consequently, when computing the atomic data with the RATIP code, we replaced the electron-nucleus and electron-electron interactions of the ionic system by the corresponding Debye-H\"uckel potential given in Equation 3 in which we modified the plasma screening effects by changing the $\lambda$ parameter value from 0 to 0.1 a.u.

\begin{table}[h]
		\caption{Values of the screening parameter $\lambda$ (in a.u.) for different plasma conditions.}
\tabcolsep7pt\begin{tabular}{lccccccc}
			\hline
                            & $n_e$ = 10$^{18}$ cm$^{-3}$ & $n_e$ = 10$^{19}$ cm$^{-3}$ & $n_e$ = 10$^{20}$ cm$^{-3}$ & $n_e$ = 10$^{21}$ cm$^{-3}$ \\
            \hline
            $T_e$ = 10$^5$ K    &    0.002           &     0.008          &      0.024         &      0.077         \\
            $T_e$ = 10$^6$ K    &    0.001           &     0.002          &      0.008         &      0.024         \\
            $T_e$ = 10$^7$ K    &    0.000           &     0.001          &      0.002         &      0.008         \\
			\hline
		\end{tabular}
\end{table}

\section{RESULTS AND DISCUSSION}

The ionization potentials (IP) and K-thresholds obtained in the present work using different values of the plasma screening parameter are given in Table 2. First of all, it is worth mentioning that the ionization potentials obtained in our work for the isolated ions are in very good agreement (within 0.1 \%) with the most accurate values published so far for Fe XXIV (IP = 2.045759(7) keV) [9] and Fe XXV (IP = 8.8281878(11) keV) [10]. Furthermore, when looking at Table 2, we note that, when going from $\lambda$ = 0 (isolated ion) to $\lambda$ = 0.1 (which corresponds to typical plasma conditions such as $T_e$ = 10$^5$ K and $n_e$ = 10$^{21}$ cm$^{-3}$), the ionization potentials of Fe XXIV and Fe XXV are reduced by 64.5 and 67.8 eV, while the corresponding K-thresholds are reduced by 65.1 and 67.8 eV, respectively. These shifts are illustrated in Figures 1 and 2. It is important to underline here that, if this effect of ionization potential depression in dense plasmas is a well-know process, this is not taken into account in a comprehensive way in current ionization balance calculations carried out in photoionized plasma modeling. As an example, in the code XSTAR [11,12], the treatment of continuum lowering for H- and He-isoelectronic sequences is based on fitting detailed cascade calculations incorporating about 10000 levels for each ion, with results applied to rates for fictitious superlevels. However, these results are only applicable up to densities of 10$^{18}$ cm$^{-3}$, which is not adequate for modeling iron lines from high density environments. In addition, the lowering of the ionization potentials can affect atomic processes such as photoionization/radiative recombination, autoionization/dielectronic recombination and collisional excitation/deexcitation.

\begin{table}[h]
		\caption{Computed ionization potentials and the K-thresholds for Fe XXIV and Fe XXV ions as a function of the plasma screening parameter $\lambda$.}
\tabcolsep7pt\begin{tabular}{lccccccc}
			\hline
			& \multicolumn{3}{c}{Ionization potential (keV)} &  \multicolumn{3}{c}{K-threshold (keV)} \\
			Ion & $\lambda = 0.00$ & $\lambda = 0.05$ & $\lambda = 0.10$ & $\lambda = 0.00$ & $\lambda = 0.05$ & $\lambda = 0.10$ \\
			\hline
            Fe XXIV  & 2.0443 & 2.0119 & 1.9798 & 8.6893 & 8.6568 & 8.6242 \\
            Fe XXV   & 8.8367 & 8.8030 & 8.7689 & 8.8367 & 8.8030 & 8.7689 \\
			\hline
		\end{tabular}
\end{table}

The wavelengths and radiative transition probabilities computed in our work for K$_{\alpha}$ lines in Fe XXIV and Fe XXV are reported in Tables 3 and 4, repectively. These results were obtained using two different values of the plasma screening parameter, i.e. $\lambda$ = 0 and 0.1 a.u. When looking at the tables, it is clear that the influence of the plasma environment on the calculated parameters is rather small in both ions, the wavelengths being shifted by about 10$^{-4}$ \AA~ while the changes observed for the radiative transition rates do not exceed 0.3 \%.

The results obtained concerning the influence of plasma environment on the atomic structure and radiative transition probabilities for all the other iron ions, from Fe I to Fe XXIII, will be published elsewhere. A first set of results obtained in the case of Ne-, Na-, Ar- and K-like iron ions will be available soon [13]. In the near future, we also intend to apply the same method as the one reported in the present paper to the study of other atomic processes such as Auger effect, radiative recombination, photoabsorption and photoionization.

\begin{table}[h]
		\caption{Comparison between the wavelengths and transition probabilities computed for the K$_{\alpha}$ lines in Fe XXIV using a plasma screening parameter $\lambda$ = 0.1 and the values obtained for an isolated ion ($\lambda$ = 0)}
\tabcolsep7pt\begin{tabular}{lcccc}
			\hline
			& \multicolumn{2}{c}{Wavelength (\AA)} &  \multicolumn{2}{c}{Transition probability (s$^{-1}$)} \\
			Transition & $\lambda = 0.0$ & $\lambda = 0.1$  & $\lambda = 0.0$ & $\lambda = 0.1$ \\
			\hline
			1s2s2p $^4$P$_{3/2}$ - 1s$^2$2s $^2$S$_{1/2}$	&	1.8718	&	1.8719	&	1.289E+13	&	1.288E+13	\\
			1s2s2p $^2$P$_{1/2}$ - 1s$^2$2s $^2$S$_{1/2}$	&	1.8612	&	1.8613	&	3.029E+14	&	3.029E+14	\\
			1s2s2p $^2$P$_{3/2}$ - 1s$^2$2s $^2$S$_{1/2}$	&	1.8585	&	1.8586	&	4.936E+14	&	4.934E+14	\\
			1s2p$^2$ $^4$P$_{1/2}$ - 1s$^2$2p $^2$P$_{1/2}$	&	1.8706	&	1.8707	&	1.728E+13	&	1.729E+13	\\
			1s2s2p $^2$P$_{1/2}$ - 1s$^2$2s $^2$S$_{1/2}$	&	1.8550	&	1.8551	&	2.009E+14	&	2.008E+14	\\
			1s2p$^2$ $^4$P$_{5/2}$ - 1s$^2$2p $^2$P$_{3/2}$	&	1.8707	&	1.8709	&	3.009E+13	&	3.007E+13	\\
			1s2p$^2$ $^2$D$_{3/2}$ - 1s$^2$2p $^2$P$_{1/2}$	&	1.8609	&	1.8610	&	3.224E+14	&	3.223E+14	\\
			1s2p$^2$ $^2$D$_{3/2}$ - 1s$^2$2p $^2$P$_{3/2}$	&	1.8655	&	1.8656	&	2.680E+13	&	2.679E+13	\\
			1s2p$^2$ $^2$P$_{1/2}$ - 1s$^2$2p $^2$P$_{1/2}$	&	1.8602	&	1.8603	&	5.548E+14	&	5.547E+14	\\
			1s2p$^2$ $^2$P$_{1/2}$ - 1s$^2$2p $^2$P$_{3/2}$	&	1.8647	&	1.8649	&	1.580E+14	&	1.576E+14	\\
			1s2p$^2$ $^2$D$_{5/2}$ - 1s$^2$2p $^2$P$_{3/2}$	&	1.8636	&	1.8637	&	2.172E+14	&	2.172E+14	\\
			1s2p$^2$ $^2$P$_{3/2}$ - 1s$^2$2p $^2$P$_{1/2}$	&	1.8552	&	1.8553	&	1.482E+13	&	1.481E+13	\\
			1s2p$^2$ $^2$P$_{3/2}$ - 1s$^2$2p $^2$P$_{3/2}$	&	1.8598	&	1.8599	&	6.425E+14	&	6.423E+14	\\
			1s2p$^2$ $^2$S$_{1/2}$ - 1s$^2$2p $^2$P$_{3/2}$	&	1.8545	&	1.8546	&	2.548E+14	&	2.548E+14	\\
			\hline
		\end{tabular}
	
\end{table}

\begin{table}[h]
		\caption{Comparison between the wavelengths and transition probabilities computed for the K$_{\alpha}$ lines in Fe XXV using a plasma screening parameter $\lambda$ = 0.1 and the values obtained for an isolated ion ($\lambda$ = 0)}
\tabcolsep7pt\begin{tabular}{lcccc}
			\hline
			& \multicolumn{2}{c}{Wavelength (\AA)} &  \multicolumn{2}{c}{Transition probability (s$^{-1}$)} \\
			Transition & $\lambda = 0.0$ & $\lambda = 0.1$  & $\lambda = 0.0$ & $\lambda = 0.1$ \\
			\hline
    		1s2p $^3$P$_1$ - 1s$^2$ $^1$S$_0$	&	1.8571	&	1.8572	&	3.555E+13	&	3.553E+13	\\
	    	1s2p $^1$P$_1$ - 1s$^2$ $^1$S$_0$	&	1.8476	&	1.8478	&	4.843E+14	&	4.842E+14	\\
			\hline
		\end{tabular}
	
\end{table}

\begin{figure}[h!]
  \centerline{\includegraphics[width=240pt,clip]{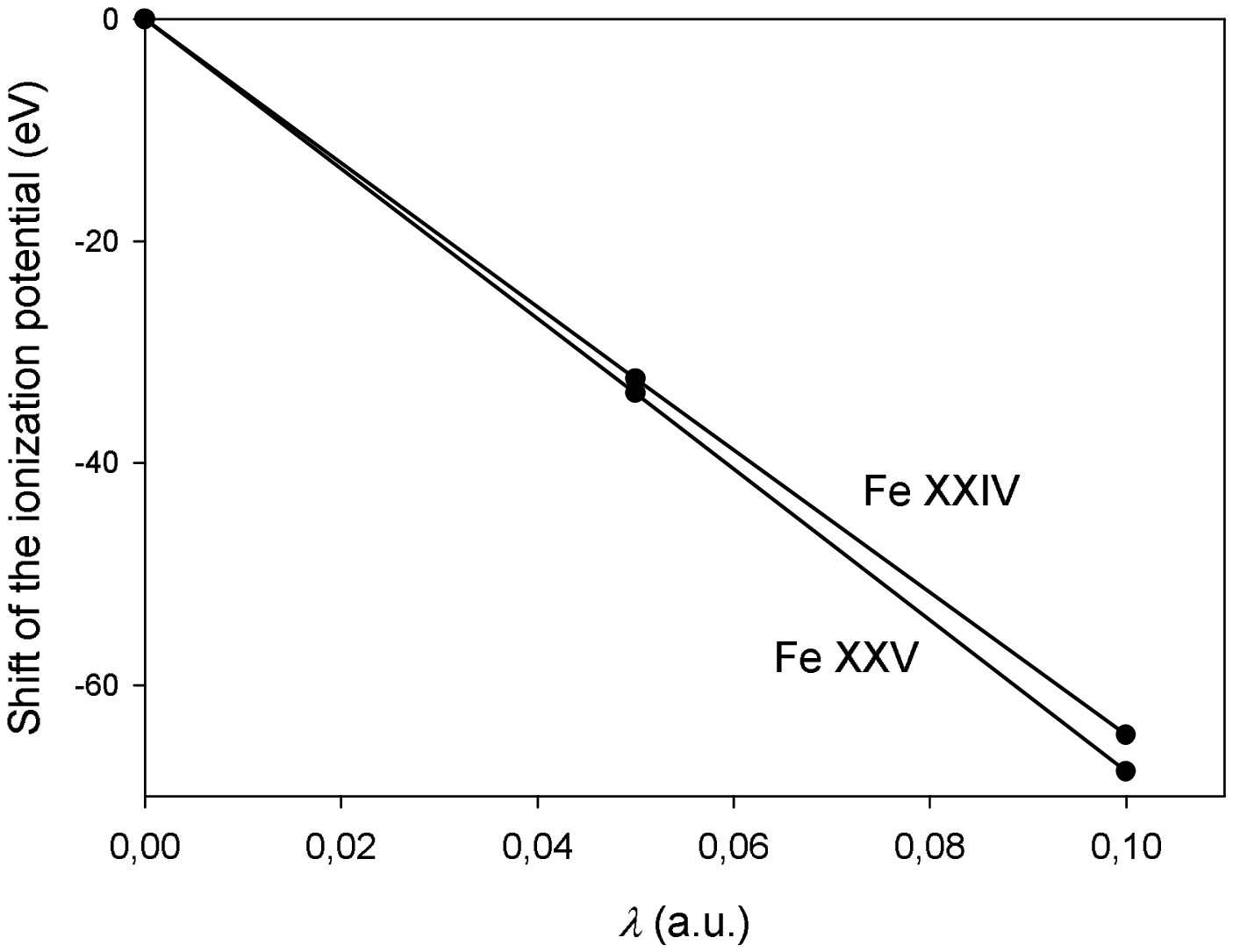}}
  \caption{Ionization potential shifts in Fe XXIV and Fe XXV as a function of the plasma screening parameter $\lambda$.}
\end{figure}

\begin{figure}[h!]
  \centerline{\includegraphics[width=240pt,clip]{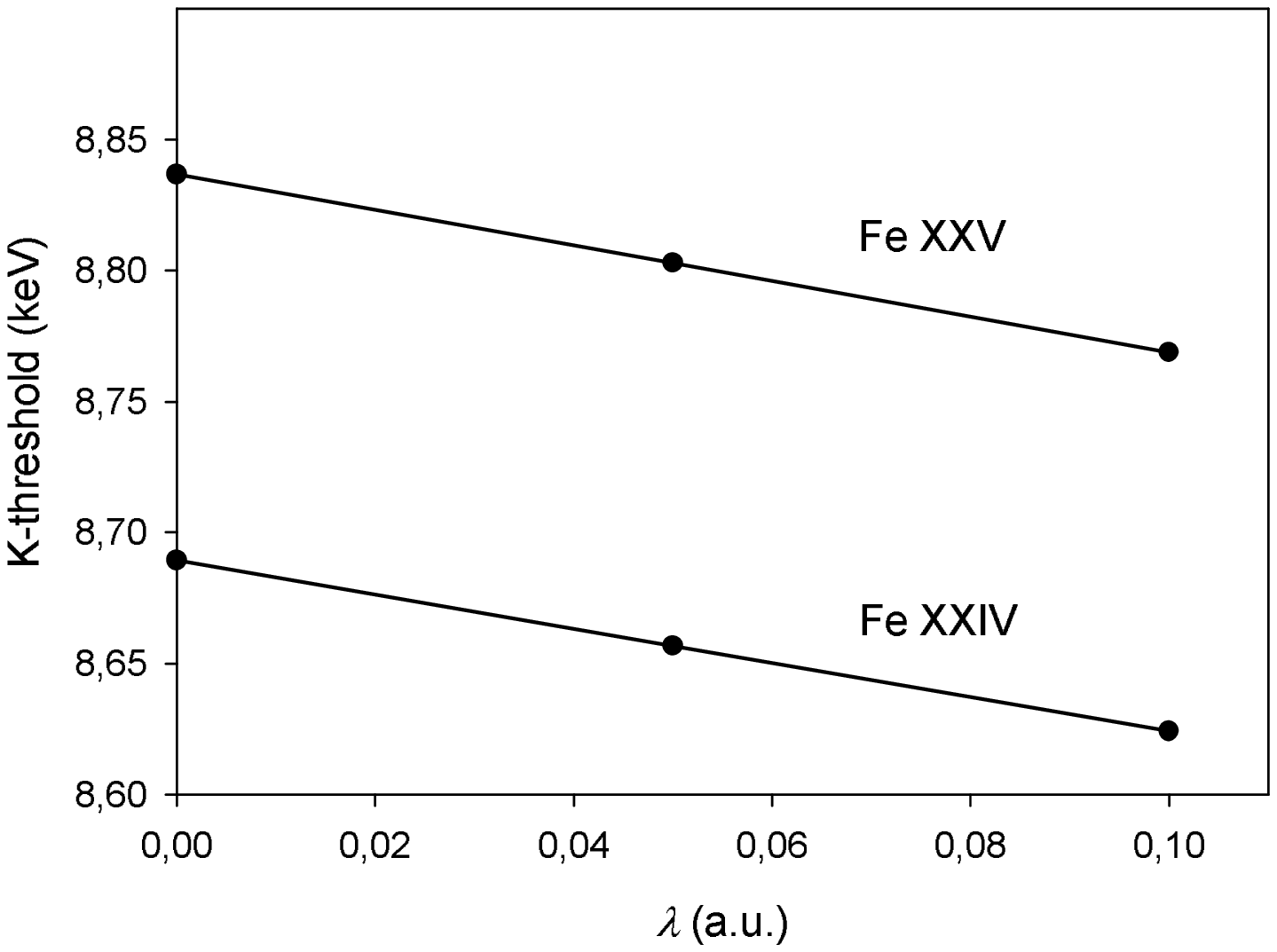}}
  \caption{Variation of the K-thresholds for Fe XXIV and Fe XXV as a function of the plasma screening parameter $\lambda$.}
\end{figure}

%
%


\newpage

\section{ACKNOWLEDGMENTS}
P.P. and P.Q. are, respectively, Research Associate and Research Director of the Belgian Fund for Scientific Research F.R.S.-FNRS. Financial support
from this organization is gratefully acknowledged.


\nocite{*}
\bibliographystyle{aipnum-cp}%
\bibliography{sample}%

\end{document}